\documentstyle[prb,aps,multicol,epsfig]{revtex}

\begin{document}
\title{Spectral diffusion on
ultralong time scales in low temperature glasses}

\author{Peter Neu, David R. Reichman, and Robert J. Silbey\\
{\it Department of Chemistry and Center for Materials Science and Engineering,\\
 Massachusetts Institute of Technology,
Cambridge, Ma.  02139, USA}\\ {\small (appears in Phys. Rev. B)}}

\maketitle
\thispagestyle{empty}

\begin{abstract} A dynamical theory is constructed to describe
spectral diffusion in glasses in the temperature range near 1 Kelvin  on long
time scales.  The theory invokes interacting tunneling centers (TLS)
which provide an excess contribution to the spectral hole width
which qualitatively accounts for the deviation from standard logarithmic
line broadening observed by Maier et al. [Phys. Rev. Lett. {\bf 76}, 2085 (1996)].
Alternative explanation schemes of the nonlogarithmic line broadening, avoiding
interacting TLS, are discussed.  We devise experimental tests which could be used to 
access the validity of the proposed theories.
\\

\end{abstract}

\begin{multicols}{2}

\section{Introduction}
At low temperatures the thermal, acoustic and optical behavior of glasses
differs significantly from that of crystalline solids. As examples note the
linear temperature dependence of the specific heat, the pronounced
absorption of sound even  below 1~K, and the anomalous broadening of
spectral holes in the homogeneous line of chromophore molecules.  It is
known now for more than twenty years that phonons cannot 
account for these observations; instead localized low-energy excitations
in the glass are needed. Since the introduction of the standard tunneling model
(STM) by Anderson, Halperin and Varma \cite{AHV} and Phillips \cite{Phil},
and the experimental observation of saturability of ultrasound by
Hunklinger et al. \cite{Hu} and Golding et al. \cite{Gold}, it is widely
accepted
that these low-energy excitations are two-level tunneling systems (TLS).
Denoting the left and right ground state of the double-well potential by
$|L\rangle$ and $|R\rangle$, respectively, the Hamiltonian reads in a
Pauli-spin representation 
\begin{equation}\label{1}
H = - {\Delta\over 2}\sigma_x - {\epsilon\over 2}\sigma_z, 
\end{equation}
where $\Delta/\hbar$ is the tunneling frequency, $\epsilon$ the asymmetry energy
and $\sigma_z = |L\rangle\langle L| - |R\rangle\langle R|$. 
In the STM the interaction between TLS is neglected, and it is assumed that 
 the tunneling parameters $\Delta$, $\epsilon$ are random variables with  distribution
\begin{equation}\label{2}
P^{(1)}(\epsilon,\Delta) d\epsilon d\Delta = {P_0\over \Delta} d\epsilon d\Delta,
 \quad \Delta \ge \Delta_{\rm min},
\end{equation}
(with $P_0 \approx 0.6 \times 10^{45}$~J$^{-1}$~m$^{-3}$ in PMMA, a polymer glass). The ensuing constant
distribution for the TLS-energy splitting
\begin{equation}\label{3}
E = \sqrt{\Delta^2 + \epsilon^2}
\end{equation}
explains in particular the linear specific heat. Including
relaxation of the TLS via the one-phonon process with rate
\begin{equation}\label{4}
R = \left({\Delta\over E}\right)^2 R_{\rm max},
\end{equation}
where $(x=E/2k_BT$)
\begin{equation}\label{5}
R_{\rm max}(E) = \alpha T^3 x^3 \coth x 
\end{equation}
and 
\begin{equation}\label{5a}
\alpha = {\gamma^2 (2k_{\rm B})^3\over 2\pi \hbar^4\varrho v^5},
\end{equation}
 the model describes most
acoustic and optical experiments in glasses satisfactorily. Here, $\gamma$
is the deformation potential energy of the TLS-phonon coupling, $\varrho$
the mass density of the glass, and $v$ the sound velocity.

Recent experiments report a systematic disagreement with the STM.
An example is the attenuation of sound below 100~mK. The STM predicts a
$T^3$-increase in contradiction with the experimental $T^{(1-2)}$-law.
\cite{Es,Enss1} Deviations from the predicted STM behavior
was also observed by Maier, Kharlamov, and Haarer
 in low temperature hole-burning experiments. \cite{Haar1,Haar2} They
performed hole-burning of a chromophore embedded in PMMA at temperatures around 1~K up to extremely 
long times (from
10~s to 10~days). They found a logarithmic time-dependence with a crossover
to an algebraic behavior after about 3~h. Though the $\log t$-behavior is in
agreement with the STM, the algebraic behavior is not. The authors could
fit their data with an  {\it ad hoc} ansatz
\begin{equation}\label{6}
P(\epsilon,\Delta) = P_0 \left[ {1\over \Delta} + {A\over \Delta^2}\right],
\qquad  A = {\rm constant},
\end{equation}
 for the TLS-parameter distribution function
and the assumption that relaxation occurs via the one-phonon process with
the rate (\ref{4})--(\ref{5a}). They motivate the distribution function
(\ref{6}) by recent publications focusing on the interaction of TLS in
glasses. \cite{K,YL,Copp,BK} Indeed, Burin and Kagan \cite{BK} have shown,
following earlier ideas of Yu and Leggett \cite{YL} that pairs 
of interacting  TLS  do provide a means of constituting a
distribution  like the second term in Eq. (\ref{6})
for certain excitations in their energy spectrum.
They called  TLS which are distributed according to the
first term in (\ref{6}) {\it primary TLS}, and those
distributed according to the second term in (\ref{6}), i.e., pairs of primary TLS,
  {\it secondary TLS}.
The nice feature of their theory is that
 the  distribution of secondary TLS [second term in Eq. (\ref{6})]
is {\it derived} 
 from primary TLS which are distributed according to the STM [first term in Eq. (\ref{6})].
In that sense, Burin and Kagan's theory stays within the framework of the STM, and includes
only excitations which have not been considered in the traditional treatment.
So far these ideas have been worked out very qualitatively, and 
 with emphasis
on experiments in the millikelvin regime. Hence, the question arises whether
they apply to the experiment of Maier et al., i.e.,  for relaxation processes at 1~K  on the 
time scale between hours and days, or whether an extension of the STM---which
would  be as  phenomenological as the STM---has to be found in order to understand this experiment.
It is the purpose of this paper to address this issue by presenting a detailed
model that includes TLS-TLS coupling in the manner proposed by Burin and Kagan.

 The paper is laid out as follows:
in Sec. II,  we analyze the experiment of Maier et al.  \cite{Haar1,Haar2} and
show the achievements and failures of Burin and Kagan's approach for experiments
in the Kelvin regime; in Sec.
III and IV, we propose a specific model that combines interacting TLS with
strong-coupling effects between TLS and phonons in the framework of the 
theory of Kassner and Silbey \cite{KS};
 in Sec. V,  we compare the predictions of the new model 
with the  hole-burning data in Ref. \CITE{Haar2},  and also discuss 
alternative  explanation schemes, which comprise an extension of the STM, and compare with equivalent
hole-burning measurements in proteins; finally, in Sec. VI, we discuss our results
and  conclude with a short summary in Sec. VII.
 The mathematical details are relegated to three
Appendices in order not to obscure the basic ideas.

\section{Hole-burning at ultralong times  and the Burin-Kagan theory}
In Ref. \CITE{Haar1,Haar2} photo-chemical hole-burning in  PMMA at 1 and
0.5~K has been performed for extremely long times $t_{\rm max} = 10^6$~s.
The authors found a $\log t$-dependence with a crossover to an algebraic
behavior  after approximately 3~h. The crossover shifts 1 order of
magnitude in time from $10^4$ to $10^3$~s, if the temperature is increased
by a factor of 2. 

A theoretical description of spectral diffusion in glasses 
was provided by Hu and Walker \cite{HW} and Black and Halperin \cite{BH}.
Reinecke \cite{Re}, and later 
Bai and Fayer \cite{BF}, extended their results  to optical experiments.
Based on this work the dependence of the hole width $\Gamma (t)$
on the waiting time $t$ is determined by 
\begin{equation}\label{8}
 P(E,r) = {P_0\over 2\sqrt{1-r}}
\left[{1\over r} + {A \over r^{3/2}E}\right], 
\end{equation}
where 
\begin{equation}\label{7}
r = R/R_{\rm max},
\end{equation}
according to  the following equation:
\begin{eqnarray}\label{ST}
\Gamma (t) =  {\pi^2\over 3\hbar} 
\langle C\rangle \int_0^\infty &dE& \,
{\rm sech}^2{E\over 2k_B T} \int_0^1 dr\,
{\epsilon\over E} \, P(E,r)\nonumber\\
&&\times  \  \left(1 - e^{-rR_{\rm max} t}\right).
\end{eqnarray}
  Here, $\langle C\rangle$ is the chromophore-TLS coupling  strength. 
The distribution $P(E,r)$ directly follows from Eq. (\ref{6}) by using
 (\ref{3})-(\ref{5}) and (\ref{7}).
With this,  we find 
\begin{eqnarray}\label{9}
\Gamma(t) &=& \Gamma(t_0)  +  {\pi^2\over 3\hbar }\, \langle C\rangle \, P_0 
\nonumber\\
&&\times \ \left[\, k_BT\,
\log{t\over t_{0}} \,+ \, A
\,\sqrt{\alpha T^3}  \left(\sqrt{t} - \sqrt{t_{0}}\right)\,\right]
\end{eqnarray}
with 
\begin{equation}\label{alpha}
\alpha = 9 \times 10^{9} \ {\rm K}^{-3} {\rm s}^{-1} \label{12}
\end{equation}
in PMMA.\cite{HS}
 To compare with the experiment, 
 we have included in $\Gamma(t_0)$ processes 
which are faster than the shortest experimental 
time $t_{0}$, for which a hole
broadening can be determined. 
In Ref. 12(a) a  fit to the experimental data
provided
\begin{eqnarray}\label{10}
A/k_B \approx 10^{-7}\ {\rm K}
\end{eqnarray}
at  1  and 0.5~K with an error of approximately  10\%.
 Furthermore, the experimental observation of spectral diffusion up to $t_{\rm max} \sim10^6$~s
suggests 
\begin{eqnarray}\label{11}
\Delta_{\rm min}/k_B \stackrel{<}{\sim}  10^{-8}\ {\rm K} 
\end{eqnarray}
according to the relation $t_{\rm max} \le  1/R_{\rm min} = 
(2k_BT/\Delta_{\rm min})^2 (1/\alpha T^3)$.

Very recently, Burin and Kagan \cite{BK} showed that weak TLS-TLS interaction
provides ultralow energy excitations. They added 
to the Hamiltonian (\ref{1}) an interaction term
\begin{equation}\label{13}
H_{\rm TLS-TLS} = - {1\over 4} \sum_{ij} J_{ij} \sigma_z^i \sigma_z^j,
\end{equation}
 where the interaction energy
\begin{equation}\label{14}
J_{ij} = {\mu_{ij}\over |{\bf r}_i - {\bf r}_j|^3}
\end{equation}
falls off with distance $|{\bf r}_i - {\bf r}_j|$ in a manner typical for a
dipole-dipole interaction. They assumed that the angular average of the
coupling is zero,
\begin{equation}\label{15}
\langle \mu_{ij}\rangle = 0,
\end{equation}
and that the TLS-TLS coupling is weak, i.e.,
\begin{equation}\label{18}
P_0 U_0 \ll 1,
\end{equation}
where $U_0$ is set by the variance 
\begin{equation}\label{16}
\langle \mu_{ij}^2\rangle = U_0^2.
\end{equation}
Such an interaction could be mediated by {\it virtual phonon exchange}
(elastic coupling) or {\it virtual photon exchange} (electrostatic
coupling) between the TLS. In the former case the energy scale of interaction
(\ref{16}) is easily found to be
\begin{equation}\label{17}
U_0 = {\gamma^2\over \pi\varrho v^2},
\end{equation}
which, indeed,  provides $P_0 U_0\ll 1$ for all 
glasses---in PMMA\cite{HS}: $U_0 \approx 1$~eV~\AA$^3$ and 
$P_0U_0 \approx 10^{-4}$.
Based on the smallness of $P_0 U_0$,
Burin and Kagan proposed that the TLS-TLS coupling 
in glasses is dominated 
 at low temperatures by an {\it up-down transition} (cf. Fig. 1).
Such an  interaction contains {\it coherence} because the up
transition of one TLS inevitably has to be followed by a down transition of
the coupled TLS. After rotating to the eigenbasis $|0_i\rangle$,
$|1_i\rangle$ of (\ref{1}) by
\begin{eqnarray}\label{19}
\sigma_z^i &=& (\epsilon_i/E_i) S^i_z - (\Delta_i/E_i) S^i_x,\\
\label{20}
\sigma_x^i &=& (\Delta_i/E_i) S^i_z + (\epsilon_i/E_i) S^i_x,
\end{eqnarray}
where $S_z^i = |0_i\rangle\langle 0_i| - |1_i\rangle\langle 1_i|$, such an
interaction is generated by the
\begin{equation}\label{21}
-{\Delta_i\Delta_j\over 4 E_i E_j} \, J_{ij} \, S_x^i S_x^j
\end{equation}
part of (\ref{13}). The eigenstates $|1\rangle \equiv |0_i,0_j\rangle$, 
$|2\rangle \equiv |0_i,1_j\rangle$, 
$|3\rangle \equiv |1_i,0_j\rangle$, 
$|4\rangle \equiv |1_i,1_j\rangle$ 
of the Hamiltonian $H_{0,ij} = - (1/2)(E_iS_z^i + E_jS_z^j)$, (cf. Fig. 2), 
become mixed due to the interaction term (\ref{21}).
 In the {\it up-down subspace} spanned by $|0_i,
1_j\rangle = |0_i\rangle \otimes |1_j\rangle$ and $|1_i, 0_j\rangle =
|1_i\rangle \otimes |0_j\rangle$
 this pair coupling can effectively be described by a
TLS-Hamiltonian of the type (\ref{1}) with $\sigma_z =
|0_i,1_j\rangle\langle 0_i, 1_j| - |1_i,0_j\rangle\langle 1_i, 0_j|$, $\sigma_x
= |0_i,1_j\rangle\langle 1_i, 0_j| + |1_i,0_j\rangle\langle 0_i, 1_j|$, and
pair asymmetry energy, pair tunneling frequency, and pair level splitting
\begin{eqnarray}
\epsilon_p &=&  E_i - E_j,\nonumber\\
\label{22}
 \Delta_p &=& J_{ij} \Delta_i\Delta_j/2E_iE_j,\\
 E_p &=&  \sqrt{\Delta_p^2 + \epsilon_p^2}.\nonumber
\end{eqnarray}
 The eigenstates for the coherently coupled pair then read
\begin{eqnarray}\label{22a}
|+\rangle &=&  \sqrt{{1+\epsilon_p/E_p\over 2}}\  |0_i,1_j\rangle
+  \sqrt{{1-\epsilon_p/E_p\over 2}}\  |1_i,0_j\rangle, \\
\label{22b}
|-\rangle &=&  \sqrt{{1+\epsilon_p/E_p\over 2}}\  |1_i,0_j\rangle
-  \sqrt{{1-\epsilon_p/E_p\over 2}}\  |0_i,1_j\rangle.
\end{eqnarray}
Clearly, the more asymmetric the pair is, i.e., the larger the energy offset
$E_i - E_j$ is, the more localized at one TLS is  the pair excitation.

Based on the distribution  (\ref{2}) of the single TLS tunneling parameters
and a uniform spatial distribution of the single TLS in the glass, Burin
and Kagan \cite{BK} derived the following distribution function for the
parameters of coherently coupled pairs
\begin{equation}\label{23}
P^{(2)}(\epsilon_p,\Delta_p) = {\pi^3\over 12} (P_0 k_B T) (P_0 U_0) {1\over
\Delta_p^2} \Theta(\Delta_p - \Delta_{p,{\rm min}} ),
\end{equation}
where $\Theta(x)$, the unit step function, is included to emphasize
that the distribution has a cutoff at small $\Delta_{p}$. Note that the density of states of 
the pairs is linearly temperature dependent.
 Coherent coupling between
pairs is destroyed if there is spontaneous decay during the up-down
transition. For primary TLS with $E\approx 2 k_B T$ this occurs with a rate
(\ref{4}), which guided Burin and Kagan to estimate the lower bound (for secondary TLS
formed from symmetric primary TLS)
\begin{equation}\label{24}
\Delta_{p,{\rm min}} = \hbar \alpha T^3.
\end{equation}
Comparing (\ref{23})  with (\ref{6}) yields 
\begin{equation}\label{24a}
A(T) = (\pi^3/12)\, (P_0 U_0)\, k_B T.
\end{equation}
This looks encouraging; 
however, after a closer examination,  there are several inconsistencies. First,
Maier et al. \cite{Haar2} could fit  their data with a  {\it temperature independent}
parameter $A$. 
 Burin and Kagan's theory
gives $A \propto T$. The 10\% variation, which Maier et al.\cite{Haar2}
 found between the 1 and 0.5~K data,
is too weak to account for the linear temperature dependence of the theory.
Second, 
putting in numbers,  we find for  PMMA 
 at 1~K:  $A(1\ {\rm K})/k_B \approx 10^{-4}$~K  and 
$\Delta_{p,{\rm min}}/k_B \approx 400$~mK, which is inconsistent with the
experimental values (\ref{10}), (\ref{11}) by several orders of magnitude.
Indeed, at 1~K, TLS with $\Delta_{\rm min}/k_B \sim 0.4$~K can never be
responsible for spectral diffusion on the time scale between $10^3$ and
$10^6$~s. However,  it is possible that very asymmetric primary TLS
are responsible for long time spectral diffusion.
For very asymmetric TLS, the estimate of Burin and Kagan (28)
would be significantly reduced.  Hence, though Burin and Kagan's  theory predicts the 
measured time-dependence
quite accurately, there
arise severe inconsistencies in orders of magnitude and the 
temperature dependence upon applying their theory at
1~K. It should be mentioned that Burin and Kagan considered only 
 the millikelvin regime, which
avoids all these problems.

\section{The Kassner-Silbey approach for primary and secondary TLS}
Though the picture developed by Burin and Kagan is very appealing, it
explains only qualitatively the hole-burning data of Maier et al.
\cite{Haar1,Haar2}. The question arises whether a microscopic calculation
can yield testable predictions based on the interacting TLS scheme of Burin
and Kagan. We carry out calculations based on strong coupling of
TLS to phonons with deformation potential  $\gamma \approx 1$~eV.  This allows
for a shifting of the rate distribution towards longer times and indeed
brings the ``calculated" value of $A$ (see (29)) into closer agreement with 
experiment.  Furthermore, the inclusion of strong coupling effects alone
gives a reasonable fit to the experiment of Maier et al. for intermediate times
(cf. Fig. 3(a)).

Let us start with the usual spin-boson Hamiltonian for an ensemble of 
TLS interacting with phonons via a strain field. In the TLS-eigenbasis
the Hamiltonian reads
\begin{eqnarray}
H &=& -{1\over 2} \sum_i E_i S_z^i + \sum_{i,q} (\bar{u}_i S_z^i - u_i S_x^i)
c_q^i
(b_q + b_{-q}^{\dagger}) \nonumber\\
\label{H1}
&& \ +\  \sum_q \hbar \omega_q b_q^{\dagger} b_q,
\end{eqnarray}
with
\begin{equation}\label{c1}
c_q^i = {\rm i} \gamma q \,\left({1\over 2\varrho V \omega_q}\right)^{1/2}\,
e^{{\rm i}{\bf q r}_i}
\end{equation}
and
\begin{equation}
u_i = \Delta_i/E_i, \qquad \bar{u}_i = \epsilon_i/E_i.
\end{equation}
According to this, the TLS become dressed with clouds of virtual phonons.
As a result the coupling between the TLS will also be changed. The
assumption traditionally made is that the dressed entities can be
considered as weakly interacting. Hence,  first order perturbation theory in
the dressed states might be sufficient at low temperatures.

Based on this picture, Kassner and Silbey \cite{KS} derived a
$S_z^iS_z^j$-interaction between TLS and  a reduction of relaxation rates
for asymmetric TLS. Compared to Eq. (\ref{4}),  they found well below the
Debye-temperature of the glass, $T\ll \Theta_{\rm D}$,
\begin{equation}\label{25}
R = \left({\Delta\over E}\right)^2 \, e^{-G (\epsilon/E)^2}\, R_{\rm max},
\end{equation}
where [cf. Eq. (\ref{5a})]
\begin{equation}\label{26}
G = (\hbar/8\pi k_B)\, \alpha \Theta_D^2.
\end{equation}
It is a peculiar feature of their approach that symmetric TLS have no
reduced rates, i.e., $R_{\rm max}$ is still given by Eq. (\ref{5}), and
have zero interaction. It is this very fact which significantly changes the
distribution function $P^{(1)}(E,r)$ of single TLS energies $E$ and
dimensionless relaxation rates $r = R/R_{\rm max}$ compared to the
STM-result (cf. first term in Eq. (\ref{8})). According to (\ref{25})
\begin{equation}\label{27}
r = \left({\Delta\over E}\right)^2 \, e^{-G (\epsilon/E)^2}
\end{equation}
(compared to the STM-result $r = (\Delta/E)^2$). With this,  Kassner and
Silbey derived the new distribution
\begin{equation}\label{28}
P^{(1)}(E,r) = {P_0\over 2r \bar{u}(r) \{1 + G[1-\bar{u}^2(r)]\}},
\end{equation}
where $\bar{u}(r) = \epsilon/E$ is the inverse function of
\begin{equation}\label{29}
r(\bar{u}) = (1-\bar{u}^2)\, e^{-G\bar{u}^2}.
\end{equation}
The  result  is a stretching of the distribution in $r$ towards
smaller values, i.e.,  $P^{(1)}(r)$ has  an extended tail for such rates.
The flaws and merits of the Kassner-Silbey way of handling strong-TLS-phonon-coupling
effects have  been discussed in Ref. \CITE{KS}. An important point is
that a $S_x^iS_x^j$-interaction between TLS, as used in Burin and
Kagan's approach, cannot be derived from this approach.

In Appendix A we have generalized Kassner and Silbey's approach to
include coherent coupling between pairs. Our procedure has been  as follows.
First, we have eliminated the diagonal ($S_z^i$) and off-diagonal ($S_x^i$)
coupling in the Hamiltonian (\ref{H1})  by two separate unitary
transformations (\ref{U1}) and (\ref{U2}). Second, instead of continuing
with the full transformed Hamiltonian, we have projected out the 
{\it one-phonon coupling part } in the dressed state basis. This has been achieved by
expanding   the generated phonon shift operators around  their mean value
up to  the first leading term in $c_q^i$.   This generates one-phonon
transition matrix elements in the four-level system of the pair and, more importantly,
Debye-Waller factors which
renormalize Burin  and Kagan's coherent coupling term  (\ref{21}) as
\begin{equation}\label{30}
- J_{ij} \Delta_i
e^{-G(\epsilon_i/E_i)^2/2}\Delta_je^{-G(\epsilon_j/E_j)^2/2} / (4 E_i E_j)
\, S_x^i S_x^j.
\end{equation}
Here,
\begin{equation}\label{J}
{1\over 4} J_{ij} = \sum_{q} {c_q^i c_{-q}^j\over\hbar \omega_q},
\end{equation}
which is equivalent to (\ref{14})-(\ref{17}). 
 The pair asymmetry energy and
tunneling amplitude read
\begin{eqnarray}\label{31E}
\epsilon_p &=&  E_i e^{-W_iG u_i^2/2} - E_j e^{-W_jG u_j^2/2},\\
\label{31}
\Delta_p &=& J_{ij}  \Delta_i e^{-G\bar{u}_i^2/2} \Delta_j e^{-G\bar{u}_j^2/2}  /2E_iE_j,
\end{eqnarray}
with $W_i = e^{-G\bar{u}_i^2/2}$. One should note that for $G\gg1$ the 
Debye-Waller factor $e^{-W_i u_i^2/2}$ is practically unity except for
nearly symmetric TLS. Since we will be mainly interested in strongly asymmetric
TLS, we will always use this simplification hereafter. In Appendix B we calculate
the pair parameter distribution function. The result is:
\begin{equation}\label{32}
P^{(2)}(\epsilon_p,\Delta_p) = P^{(2)}(\Delta_p)\, P^{(2)}(\epsilon_p),
\end{equation}
with
\begin{equation}\label{32a}
 P^{(2)}(\Delta_p)  =    {P_0\over \Delta_p^2}\,  
\, e^{-G/2} I_0^2(G/4)\,  \Theta(\Delta_p - \Delta_{p,{\rm min}}),
\end{equation}
where $I_0(z)$ is a  modified Bessel function, and
\begin{equation}
\label{32b}
 P^{(2)}(\epsilon_p)  =  f(\epsilon_{p}) + f(-\epsilon_{p}),
\end{equation}
with 
\begin{eqnarray}\label{32c}
 f(\epsilon_{p}) &=& 
{A(T)\over 1 - e^{-\beta \epsilon_p}} \, \Theta(E_{\rm max} - \epsilon_p)\ \times \\
&\times& \left\{\log\left({2\over  1 + e^{-\beta \epsilon_p}}\right)
 - \, \log\left({1 + e^{-\beta E_{\rm max}}\over 
 1 + e^{-\beta (\epsilon_p + E_{\rm max})}}\right)\right\}.\nonumber
\end{eqnarray}
The parameters $E_{\rm max},\, \Delta_{p,{\rm min}}$ are cutoffs 
which are set by the requirement of  stability  for the coupled pairs.
$E_{\rm max}$ is a cutoff  in the energy  of the {\it primary} TLS,
which, as we will see  below,  is generally a function of the temperature.
If $E_{\rm max}\gg k_B T$,
$P^{(2)}(\epsilon_p)$ is the sum of a constant  and a bell shaped curve 
centered around $\epsilon_p = 0$
with a  width of $O(k_B T)$. It  can  be approximated by
\begin{equation}\label{AB}
P^{(2)}(\epsilon_p) \approx A(T) [\log 2 + (1-\log 2) \exp(-\beta^2 \epsilon_p^2/9)]
\end{equation}
for all practical purposes.
If $E_{\rm max}\ll k_B T$, the distribution $P^{(2)}(\epsilon_p)$ becomes
independent of $\epsilon_p$,
\begin{equation}\label{AC}
P^{(2)}(\epsilon_p) \approx A(T) E_{\rm max}(T)/4k_BT.
\end{equation}
In the limit $E_{\rm max}\gg k_B T$ and  
$G, \epsilon_p/k_BT \ll 1$, we find Burin and Kagan's result
(\ref{23}). This confirms that their model is valid at ultralow temperatures
for nearly symmetric TLS with 
weak  coupling to phonons. 
In the limit $G\gg 1$,  the asymptotic
expansion of the modified Bessel function, 
$I_0(z) \approx  e^z/\sqrt{2\pi z}$,  ($z\gg 1$), provides 
a renormalization of $P_0$  by a factor $2/(\pi G)$.

\section{Stability analysis}
To determine $E_{{\rm max}}$ and $\Delta_{p,{\rm min}}$, we first 
neglect the level broadening effect of the phonons.
Then,  $E_{{\rm max}}$ scales with the glass transition temperature, and 
$\Delta_{p,{\rm min}}$ is set by the concentration
$N_p$ of pairs in the glass. To determine $N_p$, we start with estimating
the probability for 2 TLS separated by a distance ${\bf r}$ to form a pair
to $P({\bf r}) \approx P_0 U_0/(8\pi/3)|{\bf r}|^3$. Then the concentration of
pairs in a shell $[V-V/2, V + V/2]$ is given by $N_p \approx
\int_{V/2}^{3V/2} n P({\bf r}) d{\bf r}$,  where $n = P_0 k_B T$ is the number of
thermal TLS. This yields
\begin{equation}\label{33}
N_p = (1/2)(P_0 T) (P_0 U_0) \log 3,
\end{equation}
which in three dimensions is independent of the volume $V$ due to the $|{\bf
r}|^{-3}$-law of interaction. The maximum volume $V_* = (4\pi/3) |{\bf
r}_*|^3$ each of these pairs can occupy is given by $1/N_p$ which
determines the minimum interaction energy $J_{\rm min} = U_0/|{\bf
r}_*|^3$. This provides according to Eq. (\ref{31})
\begin{equation}\label{34}
\Delta_{p,{\rm min}} = {1\over 2} J_{\rm min} =
{\pi\over 3} (P_0 U_0)^2 \, k_B T
\end{equation}
as a reasonable estimation  of the lower bound. 
 Indeed, for  PMMA,  Eq. (\ref{34}) provides 
$\Delta_{p,{\rm min}}/k_B = 10^{-8}$~K which, at 1~K, corresponds
to a maximum relaxation time $\tau_{\rm max} = 10^6$ to $10^7$~s.

Let us now include decoherence effects by phonons.
We will not provide a full discussion of the relaxation dynamics in the
four-level system of the pair, but instead try  to argue more physically.
 After two unitary transformations
(\ref{U1}) and (\ref{U2}), one finds the Hamiltonian (\ref{A1}) as is 
pointed out in Appendix A.
 Here,  only the term $\propto S_x\otimes S_y$ allows relaxation
within the up-down subspace $\{|0,1\rangle, |1,0\rangle\}$. The relaxation
mechanism is a flip-flop process linked with the emission or absorption
of a phonon. The relaxation rate scales with $(\Delta_p/E_p)^2$:
\begin{equation}\label{Rpair}
R^{(p)} = r_p R^{(p)}_{\rm max},
\end{equation}
where
\begin{eqnarray}
r_p &=& \left({\Delta_p\over E_p}\right)^2 \\
R^{(p)}_{\rm max} &\approx & \alpha T^3 x_p^3 \coth x_p
\end{eqnarray}
and  $x_p = E_p/2k_B T$. 
We used that $\alpha_p \approx \alpha$, as discussed in Appendix A.
Note that there is no  Debye-Waller factor of the Kassner-Silbey type because
of the coupling of the phonons to $\sigma_y^p$ instead of to $\sigma_z^p$.
The coherent coupling is destroyed  by 
spontaneous decay during the up-down transition of primary TLS constituting a pair: 
for instance $|0,1\rangle, |1,0\rangle \to |0,0\rangle$.
Based on this argument, Burin and Kagan estimated 
 $\Delta_{p,{\rm min}} = \hbar  \alpha T^3$, where $\alpha T^3$ 
 is the decay rate of  {\it symmetric},  primary TLS with $E=2k_BT$. Clearly, 
  on a the time scale explored in the experiment of Maier et al., strongly
asymmetric TLS dominate the hole width instead of symmetric ones.  Note that 
in constructing $P^{(2)}(\epsilon_{p},\Delta_{p})$ (Eq. (42)), we integrated
over  values of the energy splittings of the primary TLS less than $E_{\rm max}(T)$, which 
have the correct initial population factors to insure the creation of stable  
pair  excitations. We now investigate which primary TLS 
are able to guarantee stability of secondary TLS at 1~K, i.e.,
we  ask whether the limit $E_{\rm max}(T) > k_BT$
or $< k_BT$ applies. We first study
if secondary TLS may be formed from primary
TLS with thermal splitting, $E \sim O(k_{B}T)$, at 1~K. 
 We need to satisfy three
conditions.  First, we require 
\begin{equation}\label{B1}
E_{p}/\hbar \ge R(E = 2k_BT),
\end{equation}
where $R(E=2k_BT) \equiv r \alpha T^3 {\le}   1/t$ 
is the relaxation rate of a primary thermal TLS
 which has not yet decayed at $t$, i.e., for which (cf.  Eq. (\ref{25}))
\begin{equation}\label{B2}
1/r \ge \alpha T^3 t \approx 10^{13} - 10^{16}
\end{equation}
with $t = 10^3 - 10^6$~s.  These two requirements guarantee that the secondary
TLS is coherent on the time scale where deviations from logarithmic spectral diffusion
is seen.
Furthermore, if  secondary TLS are responsible for the 
spectral line broadening after $10^3$~s, the pair rate must satisfy
$R^{(p)} t = 1$. This  provides the relation 
$(\Delta_p/E_p)^2 = 1/\alpha T^3 t x_p^3 \coth x_p$. 
Using this relation, multiplying Eq. (\ref{B1}) by $1/E_p$ and squaring it,
one can now easily check that Eq. (\ref{B2})  always implies Eq. (\ref{B1}).

With respect to the  stability of secondary TLS at 1~K, we conclude:
First, from the condition (\ref{B2}) we deduce the criterion  $r\ll 1$, i.e,
the primary TLS must be very asymmetric.  Second, 
upon replacing $\Delta_p = J r$ [cf. Eq. (\ref{31})] and noting
that $x_p/\sqrt{x_p^3 \coth x_p}\approx 1$ for $0\le x_p\le 1$,
 we  find the criterion 
\begin{equation}\label{B3}
{J\over 2k_B T} > \sqrt{\alpha T^3 t} \approx 10^6 - 10^8.
\end{equation}
 Hence, according to $J = U_0/d^3$ and  $U_0 \approx 10^4$~K~\AA$^3$,
the relevant TLS are separated by a distance 
 $d \stackrel{<}{\sim}1$~\AA.  Note that these estimates are highly approximate.
 In particular, we have approximated the rate at which coherence in a secondary
TLS is destroyed by the relaxation rate of a primary TLS, without consideration of the 
true rates that govern the coupled four level system.  However, we may still conclude 
that at 1~K, it is unlikely that primary TLS with thermal splitting, $E \sim O(k_BT)$,
 can form pairs, unless
they sit extremely close to each other.  This fact suggests a natural cutoff, 
$E_{\rm max} (T)$,  in the energy
splittings of the primary TLS that comprise the secondary TLS existing at 1~K.  This cutoff
is motivated by the above stability criteria, and can change the temperature dependence of 
$P^{(2)}(\epsilon_{p})$  in Eq. (\ref{32b}) and of the hole width in general. 
 We will return to this point in the 
next section.
According to this, the distribution $P^{(2)}(E_p,r_p)$ reads
\begin{eqnarray}\label{35}
P^{(2)}(E_p,r_p) =   P^{(2)}(r_p) P^{(2)}(E_p,\bar{r}_p)
\end{eqnarray}
where
\begin{equation}
 P^{(2)}(r_p) = 
{P_0 \, e^{-G/2} I_0^2(G/4)\over 2r_p^{3/2}\sqrt{1-r_p}} \, \Theta (r_p - r_{p,{\rm min}}),
\end{equation}
and
\begin{eqnarray}
&P&^{(2)}(E_p,\bar{r}_p) =
{ A(T) \over  |E_p|(1 - e^{-\bar{r}_p\beta E_p})}\ \times\\
&\times&  \left\{ \log\left({2\over  1 + e^{-\beta\bar{r}_p E_p }}\right)
 - \, \log\left({1 + e^{-\beta E_{\rm max}}\over 
 1 + e^{-\beta (\bar{r}_p E_p + E_{\rm max})}}\right)\right\} \nonumber\\
&+& \ (E_p \to -E_p),
\end{eqnarray}
with  $0< E_p \le E_{\rm max}(T)$,
 $\bar{r}_p = \sqrt{1-r}$, and $A(T)$ given by Eq. (\ref{24a}).

\section{Comparison with experiment}
We now calculate the broadening of a spectral hole in the inhomogeneous line
due to spectral diffusion induced by single TLS and pairs. According to the
standard theory \cite{HW,BH,Re,BF,Haar1}, the experimentally observed line
broadening,   $\Delta\Gamma(t) \equiv \Gamma(t) - \Gamma(t_0)$, can be written as
\begin{eqnarray}\label{36}
&\Delta& \Gamma(t)  =  {\pi^2\over 3\hbar} \langle C\rangle \int_0^\infty dE \ 
{\rm sech}^2{E\over 2k_BT}\, \times\nonumber\\&\times&  \int_{0}^{1}dr\,
\left( \bar{u}(r)P^{(1)}(E,r) + \sqrt{1-r}P^{(2)}(E,r)\right)\nonumber\\
&\times& \left(e^{-rR_{\rm max} t_0}  - e^{-rR_{\rm max} t}\right)
\end{eqnarray}
where $P^{(1)}(E,r)$ and $P^{(2)}(E,r)$ is given by (\ref{28}) and
(\ref{35}), respectively. To calculate these integrals, we replace the last factor
by the step function, which restricts  the $r$ integration to the interval 
$[1/R_{\rm max}t,1/R_{\rm max}t_0]$. This gives 
\begin{equation}\label{37}
\Delta \Gamma(t) =  \Delta \Gamma^{(1)}(t) +  \Delta \Gamma^{(2)}(t),
\end{equation}
where
\begin{equation}\label{371}
 \Delta \Gamma^{(1)}(t) = 
 {\pi^2\over 3\hbar} \langle C\rangle P_0\,k_BT\,
\log\frac{1-\bar{u}^2(1/ \alpha T^3 t_0)}{1-\bar{u}^2(1/\alpha T^3 t)}
\end{equation}
 is the contribution of the primary TLS to the spectral hole width.
To calculate the pair contribution $\Delta \Gamma^{(2)}(t)$, we  consider the limits
 $E_{\rm max}(T)\gg k_BT$ and $\ll k_BT$, separately.

\subsection{The limit $E_{\rm max}(T)\gg k_BT$}
Remember, in this limit coherently coupled pairs can be built by thermal primary TLS. 
 We find from Eq. (\ref{36})
\begin{eqnarray}\label{372}
 \Delta \Gamma^{(2)}(t) &=&
 {\pi^2\over 3\hbar}\, e^{-G/2} I_0^2(G/4)\, \langle C\rangle P_0 \, \times\nonumber\\
&\times& A(T) \,  \sqrt{\alpha T^3}\,\left(\sqrt{t} - \sqrt{t_0}\right).
\end{eqnarray}
Putting Eqs. (\ref{37})--(\ref{372}) together provides  an
  equation which  is very similar to the result (\ref{9}). However, this equation  is 
 based on  the microscopic picture  of phonon-mediated 
TLS-TLS interaction. 
 There are two regimes where the function  $\bar{u}
(1/\alpha T^3t)$ can be determined analytically:
(i) {\it short-time limit:} $\alpha T^3 t \ll e^G$ which gives 
$\bar{u}^2(1/\alpha T^3 t)
\approx {1\over 1+G} \log(\alpha T^3  t)$; (ii)
{\it longtime limit:}  $\alpha T^3 t \gg e^G$
which gives $\bar{u}^2(1/\alpha T^3 t) \approx 1 - e^G/\alpha T^3 t$. 
In Ref. 12(b), Maier and Haarer have fit their data with only the first term in (\ref{37}).
By numerical inversion of  Eq. (\ref{29}), they could find good agreement on intermediate
time scales up to $200-300$~min. We have illustrated this in Fig. 3(a). The values for the 
fit parameters $G$ and $P_0\langle C\rangle$ are 32 and $6\times 10^{-5}$, respectively.
With this value for $G$, the 
 crossover from the short-  to the longtime  behavior happens after ca. 150~min;
Eqs. (\ref{alpha}) und (\ref{26}) yield a Debye-temperature of $\Theta_D = 108$~K,
which is in reasonable agreement with literature data for PMMA.  It is in part due
to the success of the Kassner-Silbey theory at intermediate times that we have
adopted the strong coupling approach as our dynamical starting point.
 For times larger than 300~min, Maier and Haarer  attributed 
the deviations of the theory from the experimental data the contribution of  interacting TLS.
If this is true, we should find agreement between theory and experiment
when  including  the   second term in (\ref{37}). 
In Fig. 3(b) we have plotted $\Delta \Gamma (t)$,  Eqs. (\ref{37})--(\ref{372}),  together
with the experimental data of Ref. \CITE{Haar2} for 1~K (upper curve)
and 0.5~K (lower curve).
We used the same value for $G$ as in the previous plot, and 
have optimized  the TLS-TLS coupling parameter, $P_0 U_0$, and the TLS-chromophore 
coupling parameter, $P_0\langle C\rangle$,  to find best agreement for  the  0.5~K data. 
The result is $P_0U_0 = 2.5\times 10^{-6}$ and $P_0 \langle C\rangle = 4 \times 10^{-5}$. 
The upper  curve shows the prediction of the theory for the 1~K data. 
 The parameter value for $P_0 U_0$,
  which fits the 0.5~K data,  is by a factor
25--35  smaller than the literature value.\cite{HS}
 However, we see that a superposition of a $\log t$-  and a $t^{0.5}$-term can be interpreted 
as an effective $t^{0.38}$-power law on the experimental time scale
 as seen by Maier et al.  Hence, under the assumption 
that primary TLS with thermal energy splitting can form stable secondary 
TLS at 1~K, we find $\Delta \Gamma^{(2)} \propto T^{5/2}$,
giving a temperature dependence that is too strong compared
with the experimental observation which indicates  $\Delta \Gamma^{(2)} \propto T^{3/2}$,
 at least in fitting the data at the 
two temperatures, 0.5~K and 1~K. This can  clearly be
seen by the prediction of the theory for the data in Fig. 3(b).

\subsection{The limit $E_{\rm max}(T) \ll k_{B}T$}
  We have noted in the previous
section that it is very unlikely that thermal primary 
 TLS can form stable pairs at 1~K.
If we impose a minimum separation distance between primary TLS (say 5 \AA),
 then a natural energy
cutoff enters due to the stability requirements outlined in the previous section.
If we assume that this cutoff $E_{\rm max}$ satisfies the condition 
\begin{equation}
E_{\rm max} (T)\ll k_{B}T, 
\end{equation}
then we can estimate $E_{\rm max}(T)$ from the 
conditions considered in Sec. IV.  Specifically, from 
$R t\sim 1$, $R^{(p)} t\sim 1$, $x_p/\sqrt{x_p^3\coth x_p}\sim 1$, and $r = \Delta_p/J$,
\begin{equation}
\left( \frac{E_{\rm max}(T)}{2k_{B}T} \right)^{3} 
\coth\left(\frac{E_{\rm max}(T)}{2k_{B}T}\right) \sim
\frac{J_{\rm max}}{2k_{B}T \sqrt{\alpha  T^{3} t}},
\end{equation}
where $J_{\rm max} = U_{0}/d_{\rm min}^{3}$. 
 This shows that $E_{\rm max}(T) \sim T^{-1/4}$, which 
significantly alters the temperature dependence of the hole width.  If the above condition
is met, then 
\begin{equation}
P^{(2)}(E_p) \approx A(T) E_{\rm max}(T)/4k_BT  \sim  T^{-1/4},
\end{equation}
and
\begin{eqnarray}\label{G2}
\Delta \Gamma^{(2)}(t) &\approx& \frac{\pi^{2}}{6 \hbar} e^{-G/2} 
I_{0}(G/4)\, \langle C \rangle P_{0} \,(\sqrt{t} -\sqrt{t_0})\ \times\nonumber\\
&\times&\  \sqrt{\alpha T^3}\, A(T)\, \left(\frac{E_{\rm max}(T)}{2k_{B}T}\right)^2. 
\end{eqnarray}
In the regime where $E_{\rm max}(T) \ll k_{B}T$, this expression is essentially temperature
independent.  Thus, the stability requirements imply an interesting thermal breakup of
the secondary TLS. 

\section{Discussion}
 For very low temperatures ($E_{\rm max}(T) \gg k_{B}T$), the majority
of asymmetric TLS are stable even if constructed from asymmetric primary TLS that
have energy splittings on the order of $k_{B}T$, leading to the strong temperature
dependence depicted in Fig. 3(b).  We have argued that at higher temperatures, a crossover
should occur where the temperature dependence should become weaker, as secondary TLS
become less stable.  Therefore, it is qualitatively consistent with this picture
that the observed temperature dependence at $\sim 1$~K  is weaker than that shown
in Fig. 3(b).  Such arguments require further ``slowing down" of the temperature dependence.
This is indeed seen experimentally at 2~K. \cite{Haar3}  Hannig
et al. are able to fit their data with essentially the same value
of $A$ at 0.5~K and 1~K, and a value of $0.3\times A(T=0.5$~K) at 2~K. 
 Eventually, the entire TLS
picture should break down at some temperature in the range 1-10~K.
A clear test of the validity of this picture would be to see if the 
stronger temperature dependence emerges at lower temperatures.  Such
experiments are difficult to conduct at ultralow temperatures, due to slow
equilibration effects.

We may also note that the discrepancy between the 
``derived" and literature values of $P_{0}U_{0}$
can be explained by the thermal breakup of secondary TLS demanded by stability criteria.
As the temperature is raised, the primary TLS must lie closer together to form a stable
secondary TLS at long times.  The probability of finding two neighboring primary
TLS a very close distance apart is small, which effectively decreases the ``derived" value
of $P_{0}U_{0}$. Note that there is also a reduction in 
$\Delta \Gamma^{(2)}(t)$ due to the factor $(E_{\rm max}/2k_BT)^2$ (see Eq. (\ref{G2})).
 Thus, at least qualitatively, it is indeed possible that a picture based
on coupled TLS can account for all the properties currently observed in longtime
spectral diffusion experiments.

So far we have  shown the pros and cons of applying
the idea of coupled pairs to hole-burning experiments on ultralong time scales at 1~K.
Here we propose an interesting experiment which, although difficult to perform, 
would provide a conclusive  test of the coupled TLS hypothesis.
A crucial observation is that the exponent of the power law 
depends on the {\it dimensionality}  of the glassy probe. For example,
the exponent would be systematically smaller if the primary TLS were
confined to  (quasi)-two dimensions, while the interaction between
them still varies as $1/|{\bf r}|^3$. This would result in 
a  distribution function $P(\Delta_p)\propto \Delta_p^{5/3}$ (from $P(\Delta_p)\propto
P(J) = P(|{\bf r}|) |{\bf r}| (|d|{\bf r}|/dJ|)$ and $J\propto |{\bf r}|^{-3}$,
  $P(|{\bf r}|) =$~ constant) and therefore
in a $t^{1/3}$-power law. The experiment we suggest has already been performed,
albeit not not for the purpose that we discuss here and not on a
time scale up to $10^6$~s. The hole-burning experiment by Orrit et al.\cite{Orrit}
on an ionic dye in a Langmuir-Blodgett monolayer lying on 
a three-dimensional substrate is an experiment of the type 
we proposed above---the TLS dynamics is restricted to two dimensions
whereas sound waves and strain are not affected by the interface
between the amorphous layer and the bulk. Indeed, Orrit et al.\cite{Orrit}
have observed spectral diffusion which could be explained by Pack and
Fayer\cite{PF} using the standard tunneling model. From this perspective
it seems promising  to extend Orrit's experiment to longer times
and look  whether a power law weaker than the
three-dimensional result $t^{0.5}$ could be observed.

There are, however, reasons to be skeptical of the picture
we have outlined.  A number of approximations have been invoked
that render only a semiquantitative description of the experimental results.
These approximations include the reduction of the primary TLS to
an effective secondary TLS, and the use of relaxation rates for uncoupled
TLS to discuss decoherence effects for coupled secondary TLS.
These approximations, especially the first one, may not be valid at temperatures near 1~K.
Clearly, one has to think about alternative explanations.
A simple idea would be to attribute the deviation
from the standard $\log t$-behavior to  a {\it nonequilibrium} state
of those TLS  which relaxes on these long time scales. Indeed, recent
experiments by Friedrich et al. \cite{Friedrich1} have proven that
spectral diffusion in glasses under nonequilibrium conditions
 results  in a nonlogarithmic time evolution of the hole width.
However, the data in Maier's experiment
  were obtained after letting 
the sample relax  at the measurement temperature  for  a longer period
than the later data recording period. Hence, one expects that all 
relaxation processes shorter than this waiting time occur under
equilibrium conditions. 

It is interesting to compare the glass results with equivalent experiments on proteins.
 Hole-burning experiments on proteins show almost no hole broadening up to three
 hours, followed by a nonlogarithmic hole broadening regime. \cite{Friedrich2}
A new type of temperature-cycle hole-burning experiments\cite{Friedrich3}
leads to the conclusion 
that the excess broadening of the hole in the protein cannot
be interpreted in the framework of the STM. Hence, one might speculate
that in both glasses and proteins, the  interaction of TLS becomes important
at long time scales. An alternative conclusion  avoids the notion of 
interacting TLS altogether. Perhaps the energy landscape of
glasses  is not too different from that of proteins, and also shows
features of organization at high barriers.  Recall that the experiments of
Maier et al. were carried out in PMMA, which is a polymer glass.  Polymer 
glasses may be expected to have conformations similar to proteins. For example,
 such ``conformational"
dynamics may involve side chain motions.
The physical picture  is that the energy landscape 
 comprises high barriers in addition to constantly
distributed  lower barriers within each of its basins. The algebraic behavior then results 
from tunneling through those high barriers, which have to be distributed
around a ``typical'' value $V_0$. This value has to be  sufficiently high in order
that the onset of the algebraic behavior occurs only after 3~h.
We give some details on these ideas in Appendix C;
more can be found in Ref. \CITE{HN}.
 The model predicts a  temperature   and time-dependence of the hole width  with
an exponent which is slightly
weaker than $T^{3/2}$ and $t^{1/2}$, respectively, and a slowly  decreasing function of 
$T$ and $t$.  We note that if indeed specific polymer dynamics
(like side chain motion) are responsible for deviations from logarithmic spectral
diffusion, perhaps a deuterated sample may show different hole broadening
behavior. 

One is tempted to speculate that these 
 ``non-STM like'' high barriers  arise from the presence of the
chromophore in the glassy host\cite{S0}, because they have not been observed 
in sound attenuation experiments up to 100~K.\cite{Hukli,PN}
Note however, that these experiments were not performed on polymer glasses, and 
that sound attenuation experiments on PMMA and PS indeed show an additional peak above 
ca.  50~K.\cite{Pobell} Interestingly, doping a network glass with OH-impurities leads
to the same observation \cite{PN}, which conceivably supports the importants
of side chain motion in polymer glasses.  Furthermore,
in contrast to the glass, the protein probe is not
doped by a dye. Instead part of the protein is chemically changed in order
to  serve as a chromophore.  Thus, it seems that such deviations
from standard spectral diffusion behavior are not due to the inclusion
of the chromophore into the sample.

One should also note nonlogarithmic line broadening is the {\it typical}
case because a logarithmic time dependence occurs only in case
of a $1/\Delta^n$-distribution for the singular case $n=1$.
Without invoking the physical reasons for deviations from the standard $1/\Delta$ distribution
function introduced by Anderson, Halperin and Varma, and of Phillips, we may say that 
for small values of $\Delta$ (corresponding to long times), the distribution of barriers
is not really flat, but instead a smoothly varying function of the parameter $\lambda$
(see Appendix C).  That the distribution of barriers in such a model turns out to be
a  log-normal distribution shows a striking similarity to general systems exhibiting 
$1/f^{\alpha}$ noise. \cite{West}

\section{Conclusions}
In this paper we have  analyzed  the consistency of the conjecture that {\it coupled
pairs of TLS dominate spectral diffusion on ultralong
time scales}.  Because the pairs distribution $P^{(2)}(\epsilon_p,\Delta_p)$ 
is correlated with  the distribution of the primary ``STM-like'' TLS,
we have in a sense pushed the STM as far as possible by looking at
 these new low energy excitations.
We believe that this is an important step, which has to be done {\it before} trying
to find another extension of the STM for every new experiment. 
The questions was whether they also apply
to the 1~K regime, and, in particular, whether they can provide an explanation of the longtime
hole-burning experiment of Maier et al. We find that a picture based on the idea of interacting
tunneling systems seems consistent with the experimental data, although we are
unable to fit certain aspects of the experiment, such as the temperature dependence,
 quantitatively.
Also, alternative explanations have been presented.  These models are, at the moment,
 at least as speculative as the scenario of coherently coupled pairs.
  For this reason, we have discussed some possible
experimental tests of the theoretical models we have presented.
More theory and experiments have to be done to finally evaluate the role of
 coupled TLS in glasses,
and to understand the origin of nonlogarithmic hole broadening in glasses.

\section*{Acknowledgements}
This work has been supported in part by the NSF and the Alexander von Humboldt
foundation. We would like to thank  A. L. Burin, D. Haarer, 
Yu. Kagan, B. M. Kharlamov,  L. S. Levitov, H. Maier, and R. Wunderlich 
for helpful discussions.

\setcounter{equation}{0}
\appendix

\renewcommand{\theequation}{\Alph{section}.\arabic{equation}}
\section{}
In this Appendix we generalize Kassner and Silbey's approach to include
coherent coupling between pairs.  For this goal we first apply the unitary transformation
\begin{equation}\label{U1}
U_1 = \exp\left[\sum_{i,q} (2\bar{u}_i  c_q^i/\hbar\omega_q)
(b_q - b_{-q}^{\dagger})  S_z^i/2 \right] \equiv \prod_i e^{\varphi_iS_z^i/2}  
\end{equation}
to the Hamiltonian (\ref{H1}) which eliminates the diagonal coupling of
phonons to $S_z^i$. The transformed Hamiltonian $H\rightarrow U_1^{-1}HU_1$
reads with $S^j_\pm = (1/2)(S^j_x \pm  {\rm i} S^j_y)$: \cite{KS}
\begin{eqnarray}\label{H2}
H &=& -{1\over 2} \sum_i E_i S_z^i + \sum_q \hbar\omega_q b_q^{\dagger} b_q \\
&-& \sum_{i,q}  u_i  c_q^i
(b_q + b_{-q}^{\dagger})\, (B_-^i S_+^i + B_+^i S_-^i) \nonumber \\
&-& {1\over 4} \sum_{i,j} J_{ij}\, 
\left( \bar{u}_i \bar{u}_j  S_z^i S_z^j  - 2  u_i \bar{u}_j
(B_-^i S_+^i + B_+^i S_-^i)   S_z^j\right), \nonumber
\end{eqnarray}
where
\begin{equation}
{1\over 4} J_{ij} = \sum_q{c_q^i c_{-q}^j\over \hbar\omega_q},
\end{equation}
and 
\begin{equation}\label{PS}
B_\pm^i = \exp\left [\pm \sum_q (2 \bar{u}_i c_q^i /\hbar \omega_q) (b_q -
b_{-q}^{\dagger})\right] \equiv e^{\pm \varphi_i}.
\end{equation} 
The $B^i_\pm$ are the usual phonon shift operators.
Instead of proceeding with the full transformed
Hamiltonian, we project out the {\it one-phonon fluctuations}
around the shifted harmonic oscillators coordinates. 
This is achieved  by expanding  $B^i_\pm \equiv \langle B^i_\pm\rangle
+ (B^i_\pm - \langle B^i_\pm\rangle) = \langle B^i_\pm\rangle \pm \varphi_i$,
where  
\begin{equation}\label{B}
\langle B_\pm^i\rangle =
e^{-G(\epsilon_i/E_i)^2/2},
\end{equation}
and neglecting two-phonon terms.
Applying Fermi's Golden Rule to the remaining  one-phonon 
term directly yields the rate
(\ref{25}).  A coherent coupling between pairs is generated by eliminating
this term by a second unitary transformation
\begin{eqnarray}
\label{U2}
U_2 &=& \exp\left[- 
\sum_{i,q} \left(2u_i\langle B^i_\pm\rangle  c_q^i/\hbar\omega_q\right)
(b_q - b_{-q}^{\dagger}) S_x^i/2\right] \nonumber\\
&\equiv& \prod_i e^{- \phi_i S_x^i/2}.
\end{eqnarray}
This yields:
\begin{eqnarray}\label{A1q}
H &=& -{1\over 2} \sum_i E_i \widetilde{S}_z^i 
+ \sum_q \hbar\omega_q b_q^{\dagger} b_q
  \nonumber\\
&-& {1\over 4} \sum_{i,j} J_{ij}\, 
\left\{ \bar{u}_i \bar{u}_j  \widetilde{S}_z^i \widetilde{S}_z^j  +
 u_i \langle B^i_\pm\rangle u_j\langle B^j_\pm\rangle\, S_x^iS_x^j \right.\nonumber\\
 &-&  \left.2  u_i \bar{u}_j 
\Big(\langle B^i_\pm\rangle S_x^i - {\rm i}\varphi_i\widetilde{S}_y^i\Big)
   \widetilde{S}_z^j\right\},
\end{eqnarray}
where
\begin{eqnarray}\label{A2}
\widetilde{S}_z^j &=& S_z^j \cosh \phi_j  -  {\rm i}  S_y^j \sinh\phi_j,\\
\widetilde{S}_y^j &=& S_y^j \cosh \phi_j  +  {\rm i}  S_z^j \sinh \phi_j.
\end{eqnarray}
Expanding again around the one-phonon fluctuations  around the shifted 
harmonic oscillator coordinate by replacing 
$\widetilde{S}_z^j \approx  S_z^j \langle D_\pm^j\rangle  -  {\rm i} \phi_j S_y^j$ and
$\widetilde{S}_y^j \approx  S_y^j \langle D_\pm^j\rangle  +  {\rm i} \phi_j S_z^j$, where
\begin{equation}\label{A3}
 \langle D_\pm^j\rangle = e^{- G  (\Delta_j \langle B_\pm^j\rangle/E_j)^2 /2},
\end{equation}
one finds
\begin{eqnarray}\label{A1}
H &=& -{1\over 2} \sum_i E_i \langle D_\pm^i\rangle S_z^i 
+ \sum_q \hbar\omega_q b_q^{\dagger} b_q + {{\rm i}\over 2}\sum_i E_i \phi_i S_y^i
  \nonumber\\
&-& {1\over 4} \sum_{i,j} J_{ij}\, 
\left\{ \bar{u}_i \langle D_\pm^i\rangle \bar{u}_j \langle D_\pm^j\rangle
 S_z^i S_z^j \right.\nonumber\\  &+&\left.
 u_i \langle B^i_\pm\rangle u_j\langle B^j_\pm\rangle\, S_x^iS_x^j
- 2  u_i  \langle B^i_\pm\rangle \bar{u}_j \langle D_\pm^j\rangle S_x^i S_z^j\right.
\nonumber\\
&+& \left. 2\, {\rm i}\, u_i  \langle B^i_\pm\rangle \bar{u}_j \phi_j S_x^i S_y^j
 + 2\,  {\rm i}\, 
 u_i  \langle D^i_\pm\rangle \bar{u}_j \langle D^j_\pm\rangle \varphi_j S_y^i S_z^j
\right.\nonumber\\
 &-&\left. {\rm i} \, \bar{u}_i \bar{u}_j ( \langle D_\pm^j\rangle \phi_i S_y^i S_z^j + 
\langle D_\pm^i\rangle \phi_j S_z^i S_y^j)\right\}.\nonumber\\
\end{eqnarray}
The only operators which act nontrivially in the up-down
 subspace $\{|0,1\rangle, |1,0\rangle\}$
are $S_z\otimes {\bf 1}$, ${\bf 1} \otimes S_z$,
 $S_x\otimes S_x$,  $S_x\otimes S_y$, and  $S_y\otimes S_x$. If we define new
pseudo-spin operators for pairs $(i,j)$
\begin{eqnarray}\label{psspin}
\sigma_x^p &=& |0_i,1_j\rangle\langle 1_i,0_j| + |1_i,0_j\rangle\langle 0_i,1_j|\ ,\\
\sigma_y^p &=& - {\rm i}
 |0_i,1_j\rangle\langle 1_i,0_j| +  {\rm i} |1_i,0_j\rangle\langle 0_i,1_j|,\\
\sigma_z^p &=& |0_i,1_j\rangle\langle 0_i,1_j| - |1_i,0_j\rangle\langle 1_i,0_j|,
\end{eqnarray}
we can project the quoted operators onto the up-down subspace, which yields
  $S_z\otimes {\bf 1} \to  \sigma_z^p$, ${\bf 1} \otimes S_z\to -\sigma_z^p$,
 $S_x\otimes S_x \to \sigma_x^p$,  $S_x\otimes S_y \to -\sigma_y^p$, and  
$S_y\otimes S_x \to \sigma_y^p$. If we project the Hamiltonian (\ref{A1}) accordingly,
we find for each pair 
\begin{eqnarray}\label{A4}
H_{{\rm pair}} &=&  - {\Delta_p\over 2}\, \sigma_x^p - 
                          {\epsilon_p\over 2}\, \sigma_z^p
+  {\rm i} \Delta_p \sigma_y^p \sum_q {2c_q^p\over \hbar\omega_q} 
(b_q - b_{-q}^{\dagger})
\nonumber\\
&&\ +\  \sum_q \hbar\omega_q b_q^{\dagger} b_q ,
\end{eqnarray}
with 
\begin{eqnarray}
\Delta_p &=& J_{ij}\ {\Delta_i\Delta_j\over 2E_iE_j}
 \,\langle B_\pm^i\rangle \langle B_\pm^j\rangle, \\
\epsilon_p &=& E_i \langle D_\pm^i\rangle - E_j \langle D_\pm^j\rangle,\\
c_q^p &=&  c_q^i \,(\epsilon_i/E_i) - c_q^j (\epsilon_j/E_j).
\end{eqnarray}
Applying Fermi's Golden Rule to the interaction term in (\ref{A4})  provides
 a one-phonon relaxation rate 
\begin{equation} \label{RP}
R_p =  \alpha \Delta_p^2 E_p/(2k_B)^3 \,  \coth(E_p/2k_BT).
\end{equation}
Here, we have used that the oscillating term in $c_q^p c_q^{p*}$ is small
so that pairs and single TLS relax approximately with the 
same coupling parameter $\alpha$ provided that  $\epsilon_i/E_i \approx 1$.
Note that because of the coupling of 
phonons to $\sigma_y^p$ (instead of $\sigma_z^p$), there is no diagonal
coupling term $\propto S_z^p$ as in (\ref{H1}) after rotation 
to the pair-TLS eigenbasis.


\section{}
In this Appendix we calculate the distribution function for the pair parameters
$\Delta_p$ and $\epsilon_p$ of Eqs. (\ref{31E}) and  (\ref{31}). We note that the 
single-TLS distribution function for $u = \Delta/E$ and $E= \sqrt{\Delta^2 + \epsilon^2}$
reads $P(E,u) = P_0/(u \sqrt{1-u^2})$. The distribution function for the pair parameter
is then given by
\begin{eqnarray}
P^{(2)}(&\epsilon_p&,\Delta_p) = 
  \int {dE_1\over 1+e^{-\beta E_1}}\int {dE_2\over 1+e^{\beta E_2}}\nonumber\\
&\times& \int_0^1 du_1\int_0^1 du_2 \int dJ
 \,  P(E_1,u_1) P(E_2,u_2) P(J)\nonumber\\
 &\times& {1\over 2}\left\{ \delta(\epsilon_p - E_1 + E_2) + \delta(\epsilon_p + E_1 - E_2)
\right\}\nonumber\\
&\times& 
\delta\left(\Delta_p - {\textstyle{{1\over 2}}} 
J u_1 u_2 e^{-G(1-u_1^2)/2} e^{-G(1-u_2^2)/2}\right).
\end{eqnarray}
The factors $(1 + e^{\pm\beta E})^{-1}$ account for the thermal occupation of the 
primary TLS. If the TLS are homogeneously distributed in the glass 
and interact via a dipolar coupling,
 $J = U_0/|{\bf r}|^3$,  the distribution  function of $J$ reads for a three-dimensional probe
\begin{equation}
P(J) = {4\pi\over 3} {U_0\over J^2}.
\end{equation}
Using now that 
\begin{eqnarray}
\int_0^1 du {e^{-u^2G/2}\over \sqrt{1 - u^2}} &=& {\pi\over 2} e^{-G/4} I_0(G/4),
\end{eqnarray}
where $I_0(z)$ is a modified Bessel function, and
\begin{eqnarray}
\int_{y}^{1} {dx\over (1+x)(1+\mu x)} = 
{1\over 1-\mu}\left[\log {2\over 1+y} \ - \ \log{1+\mu\over 1+\mu y}\right],\nonumber
\end{eqnarray}
with $x = e^{-\beta E}$, $y = e^{-\beta E_{\rm max}}$,
 and $\mu =  e^{-\beta\epsilon_p}$, 
one easily finds Eqs. (\ref{32})--(\ref{32c}).

\section{}

In this Appendix we show how a modified tunneling model which comprises 
 an additional  Gaussian distribution in the tunneling parameter
$\lambda$ centered around the mean value $\lambda_0 \gg 1$ can provide
an algebraic line broadening. The new distribution reads with 
$\Delta = \hbar\omega_0  \, e^{-\lambda}$:
\begin{equation}\label{B8}
P(\epsilon,\lambda) = P_0 + {P_0 \widetilde{A} \over \sqrt{2\pi\sigma^2}}\,
e^{-(\lambda - \lambda_0)^2/2\sigma^2},
\end{equation}
where $\widetilde{A}$ is a dimensionless constant, and $\lambda \ge \lambda_{\rm min}(E)$,
$\lambda_{\rm min}(E) = \log(\hbar\omega_0/E)$.
A similar  model 
has been used previously  by Jankowiak and Small 
\cite{S0} and by  Zimdars and  Fayer \cite{Fay} to discuss 3-pulse photon echo data.\cite{Lit,Wiers}
A combination of both terms is needed  in order that the onset of the
algebraic line broadening does not occur too early.
In the following it will turn out sufficient to reduce the number
of parameters by  setting
\begin{equation}
\sigma^2 \equiv \lambda_0.
\end{equation}
The ensuing distribution in $\Delta$ then reads
\begin{equation}\label{B9}
P(\epsilon,\Delta) = P_0 \left[{1\over \Delta} + { B\,
(\hbar\omega_0)^{1 - (1/2\lambda_0) \log(\hbar\omega_0/\Delta)}\over
\Delta^{2 - (1/2\lambda_0) \log(\hbar\omega_0/\Delta)}} \right],
\end{equation}
where $B= \widetilde{A} e^{-\lambda_0/2}/
\sqrt{2\pi\lambda_0}$. This distribution has to be compared with (\ref{6}).
Defining the maximum relaxation rate at given $E$ by 
$R\equiv e^{-2(\lambda - \lambda_{\rm min}(E))} R_{\rm max}(E)$,
the distribution function in relaxation rates, $R$,
 and TLS-energies, $E$, reads
\begin{equation} \label{B10}
(\epsilon/E)\, P(E,R)  =
 {P_0 \over 2 } \left[ {1\over R} +
{B\,  R_{\rm max}^{\nu(R)} \over
 R^{1+\nu(R)} }
\right],
\end{equation}
where
\begin{equation}\label{B11}
\nu(R) = {1\over 2} - {1\over 8\lambda_0}\, \log (R_{\rm max}/R).
\end{equation}
From this expression,  it is obvious that an algebraic
line broadening $\Delta\Gamma(t) \propto t^{\mu}$ with an exponent $\mu <
0.5$ occurs for $\lambda_0 \gg \log(R_{\rm max}(T) t)$. The exact calculation reveals
\begin{eqnarray}\label{B12}
\Delta\Gamma (t) &=& {\pi^2\over 3\hbar} P_0 \langle C\rangle\, k_B T \,
\Big[ \log(t/t_0) \nonumber\\
&+&   \widetilde{A} \left\{ {\rm erfc}\left({\lambda_0 -
(1/2)\log(KT  t) \over \sqrt{2\lambda_0}}\right) \right.\nonumber\\
&&\left. \ - \
{\rm erfc}\left({\lambda_0 - (1/2)\log(KT t_0) \over
\sqrt{2\lambda_0}}\right)
\right\}\Big],
\end{eqnarray}
where erfc$(x)$ is the complementary error function, and $K \equiv (\hbar\omega_0/k_B)^2\alpha$.
  With $\hbar\omega_0/k_B = O(1$~K), one finds 
$K \approx 10^{10}$~K$^{-1}$s$^{-1}$.
 For $\lambda_0 \gg
\log(KT t)$ this gives an algebraic line growth 
\begin{equation}
\Delta\Gamma (t) \propto 
BT(KT t)^{(1/2) - (1/8\lambda_0)\log (KT  t)}.
\end{equation}
More details and a comparison to experiment can be found in Ref. \CITE{HN}.

%
%

\newpage

\begin{center}
{\bf FIGURE CAPTIONS}
\end{center}
\begin{itemize}
\item[FIG. 1:] Coherent up-down coupling between two TLS.
\item[FIG. 2:] Energy levels and eigenstates of the Hamiltonian
 $H_{0,ij} = - (1/2)(E_iS_z^i + E_jS_z^j)$, $(E_i> E_j)$.
Framed are the up-down states which build a basis for the secondary TLS.
\item[FIG. 3:] (a) Hole broadening in PMMA at 1~K  (upper curve)
and 0.5~K (lower curve) compared with the Kassner-Silbey theory for single TLS.
 The experimental data are from Ref. \CITE{Haar2}.
The two solid lines are $\Delta\Gamma^{(1)}(t)$, Eq. (\ref{371})  for 1 (upper solid curve)
and 0.5~K (lower solid curve) with  $G = 32$ ($\Theta_D = 108$~K)
 and $P_0\langle C\rangle = 6 \times 10^{-5}$.
(b) Same figure as above but now with the contribution of both the  single and pair TLS
[logarithmic and algebraic part of  Eq. (\ref{37})]. The parameters
$P_0\langle C\rangle = 4 \times 10^{-5}$ and $P_0U_0 = 2.5 \times 10^{-6}$
 have been optimized to find best agreement with the 0.5~K data; $G = 32$  has been kept fixed.
Comparison with the 1~K  data clearly shows that the temperature
dependence as predicted by Eqs. (\ref{24a}) and (\ref{37}) is too strong.
\end{itemize}

\end{multicols}


\vspace{1cm}

 \begin{LARGE} \underline{FIG. 1: Neu et al.} \end{LARGE}
\begin{center}
\epsfig{file=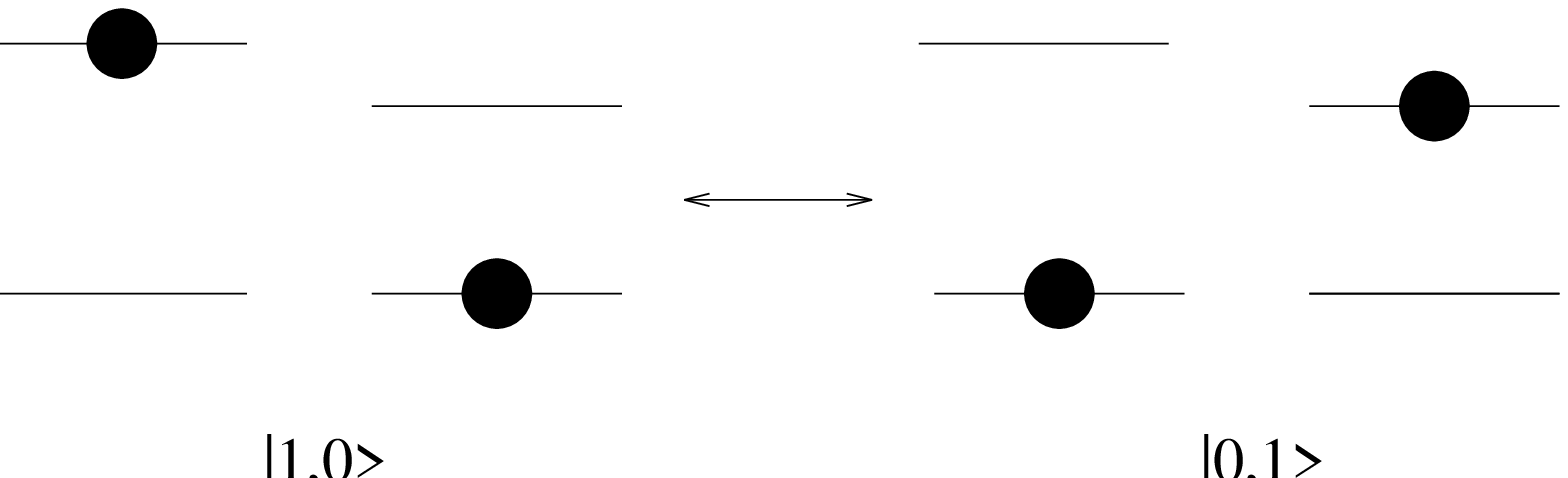}
\end{center}

\vspace{3cm}

\begin{picture}(500,300)(50,20)
\thicklines
\put(200,20){\line(1,0){100}}
\put(200,120){\line(1,0){100}}
\put(200,150){\line(1,0){100}}
\put(200,250){\line(1,0){100}}
\put(50,300){\mbox{\underline{\LARGE{FIG. 2: Neu et al.}}}}
\put(120,17){\mbox{\LARGE{$|0_i,0_j\rangle$}}}
\put(120,247){\mbox{\LARGE{$|1_i,1_j\rangle$}}}
\put(330,17){\mbox{\LARGE{$-{1\over 2}(E_i + E_j)$}}}
\put(330,247){\mbox{\LARGE{${1\over 2}(E_i + E_j)$}}}

\put(120,117){\mbox{\LARGE{$|0_i,1_j\rangle$}}}
\put(120,147){\mbox{\LARGE{$|1_i,0_j\rangle$}}}
\put(330,117){\mbox{\LARGE{$-{1\over 2}(E_i - E_j)$}}}
\put(330,147){\mbox{\LARGE{$+{1\over 2}(E_i - E_j)$}}}
\put(110,88){\dashbox{8}(340,94)}
\end{picture}
\vspace{3cm}

\begin{center}
\epsfig{file=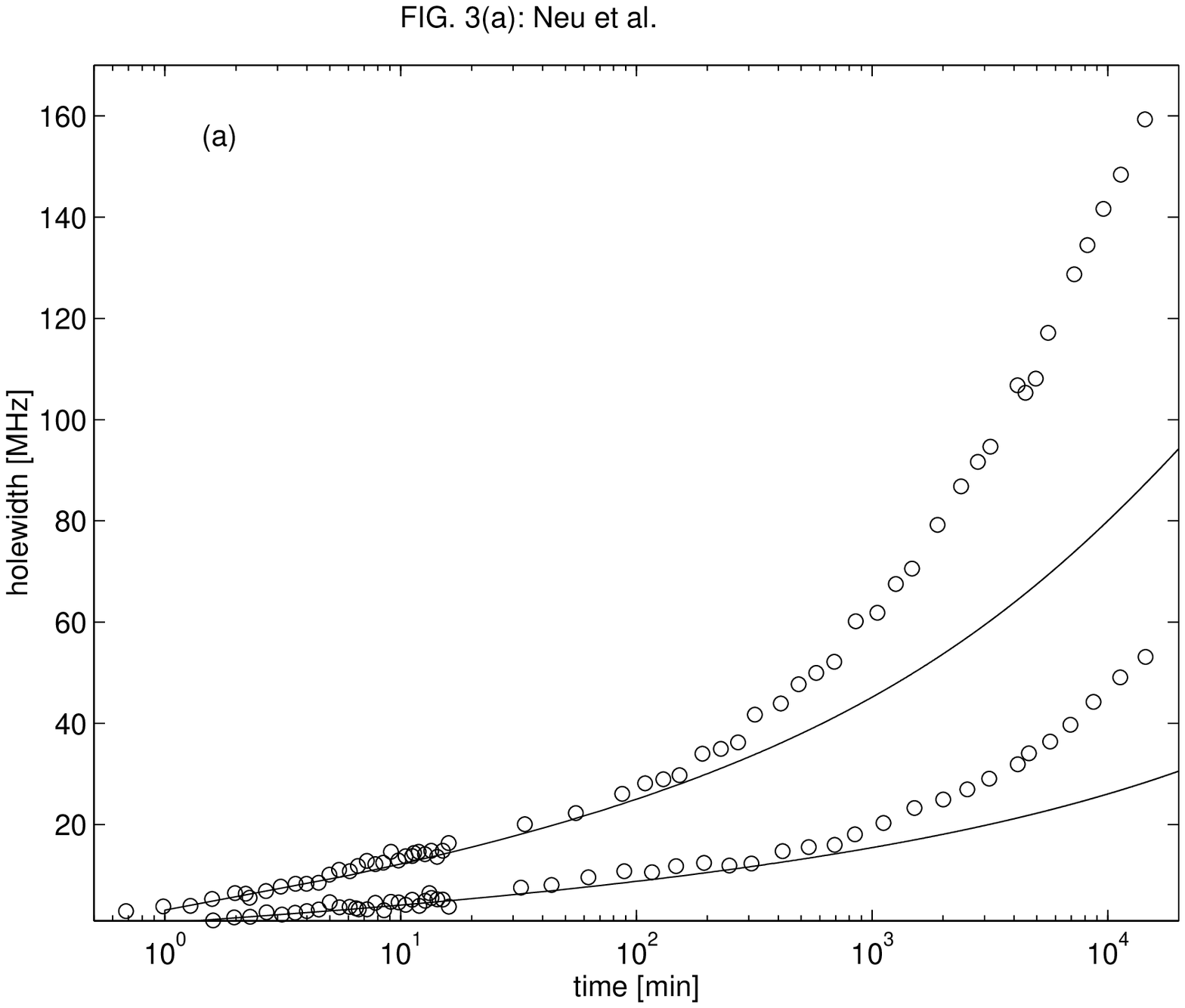}
\end{center}

\vspace{3cm}

\begin{center}
\epsfig{file=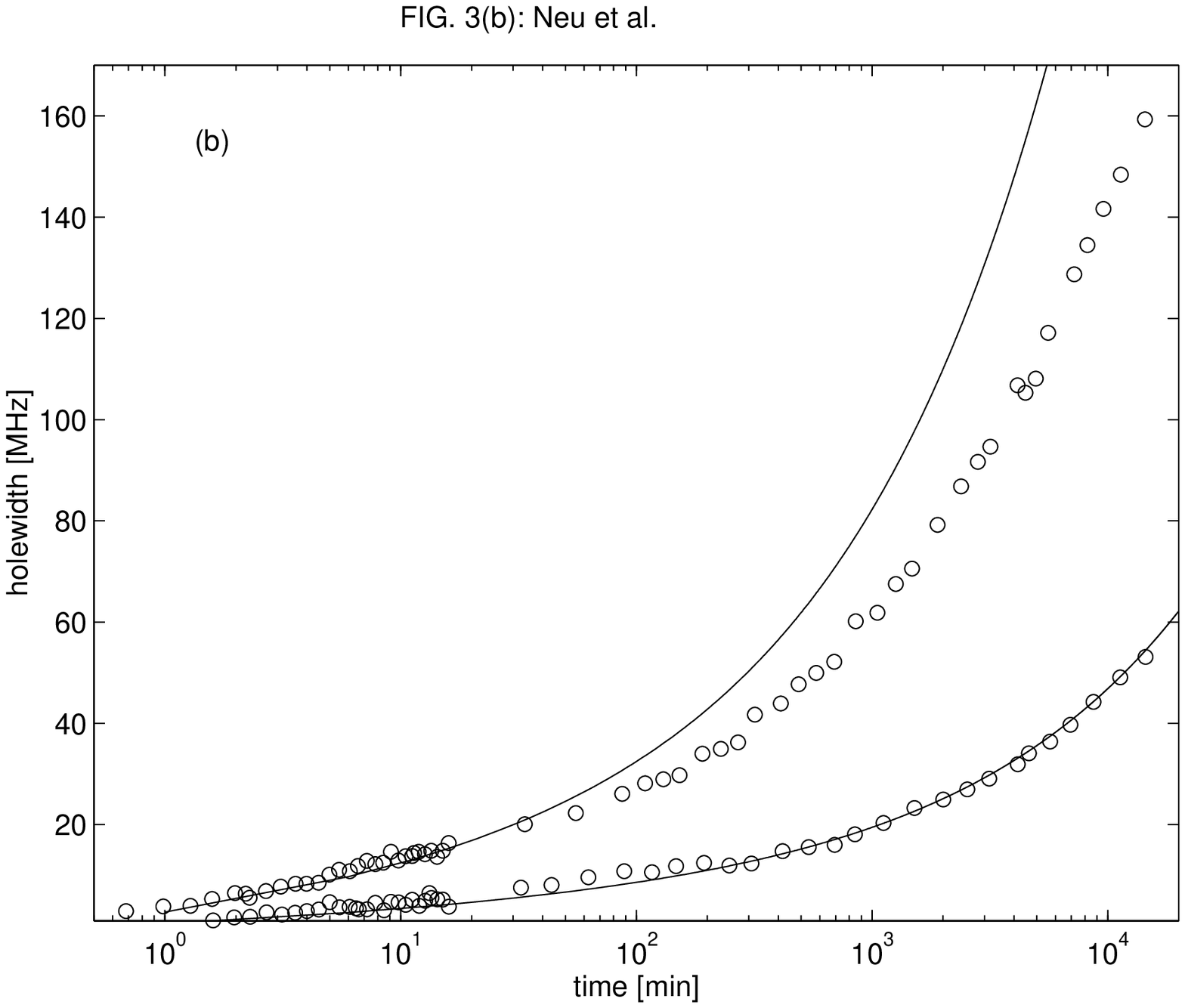}
\end{center}

\end{document}